\documentclass[12pt]{article}
\pdfoutput=1
\usepackage{jheppub}

\usepackage{amsmath,bbm,array,amsfonts,graphicx,wrapfig,lscape,float,slashbox,multirow,longtable,rotating,subfigure,epstopdf}

\newcommand{\be}{\begin{equation}}
\newcommand{\ee}{\end{equation}}
\newcommand{\beq}{\begin{equation}}
\newcommand{\beql}[1]{\begin{equation}\label{#1}}
\newcommand{\eeq}{\end{equation}}
\newcommand{\ba}{\begin{array}}
\newcommand{\ea}{\end{array}}
\newcommand{\bea}{\begin{eqnarray}}
\newcommand{\beal}[1]{\begin{eqnarray}\label{#1}}
\newcommand{\eea}{\end{eqnarray}}
\newcommand{\ben}{\begin{enumerate}}
\newcommand{\een}{\end{enumerate}}
\newcommand{\bean}{\begin{eqnarray*}}
\newcommand{\eean}{\end{eqnarray*}}

\newcommand{\nn}{\nonumber}

\newcommand{\fref}[1]{Figure \ref{#1}}
\newcommand{\btab}[1]{\begin{tabular}{#1}}
\newcommand{\etab}{\end{tabular}}

\newcommand{\comment}[1]{}

\newcommand{\IC}{\mathbb{C}}

\newcommand{\qed}{\nobreak \ifvmode \relax \else
      \ifdim\lastskip<1.5em \hskip-\lastskip
      \hskip1.5em plus0em minus0.5em \fi \nobreak
      \vrule height0.75em width0.5em depth0.25em\fi}

\def\beqa{\begin{eqnarray}}
\def\eeqa{\end{eqnarray}}
\def\NN{{\cal N}}

\newcolumntype{C}[1]{>{\centering\arraybackslash}m{#1}}

\newcommand{\IS}{{\bf S}}
\newcommand{\IZ}{{\mathbb{Z}}}
\def\II{\relax{\rm I\kern-.18em I}}

\def\makeatletter{\catcode`\@=11}
\makeatletter
\def\mathbox#1{\hbox{$\m@th#1$}}%
\def\math@ccstyles#1#2#3#4#5#6#7{{\leavevmode
     \setbox0\mathbox{#6#7}%
     \setbox2\mathbox{#4#5}%
     \dimen@ #3%
     \baselineskip\z@\lineskiplimit#1\lineskip\z@
     \vbox{\ialign{##\crcr
            \hfil \kern #2\box2 \hfil\crcr
            \noalign{\kern\dimen@}%
            \hfil\box0\hfil\crcr}}}}
\def\mathaccstyles{\math@ccstyles\maxdimen}
\def\maththroughstyles{\math@ccstyles{-\maxdimen}}
\def\unity%
{\maththroughstyles{.45\ht0}\z@\displaystyle {\mathchar"006C}\displaystyle 1}

\newcommand{\drawsquare}[2]{\hbox{%
\rule{#2pt}{#1pt}\hskip-#2pt
\rule{#1pt}{#2pt}\hskip-#1pt
\rule[#1pt]{#1pt}{#2pt}}\rule[#1pt]{#2pt}{#2pt}\hskip-#2pt
\rule{#2pt}{#1pt}}

\newcommand{\fund}{~\raisebox{-.5pt}{\drawsquare{6.5}{0.4}}~}
\newcommand{\antifund}{~\overline{\raisebox{-.5pt}{\drawsquare{6.5}{0.4}}}~}

\newcommand{\asymm}{~\raisebox{-3.5pt}{\drawsquare{6.5}{0.4}}\hskip-6.9pt%
        \raisebox{3pt}{\drawsquare{6.5}{0.4}}~}

\newcommand{\antisymm}{~\overline{\raisebox{-.5pt}{\drawsquare{6.5}{0.4}}\hskip-0.4pt%
        \raisebox{-.5pt}{\drawsquare{6.5}{0.4}}}~}

\def\IC{{\bf C}}
\def\IS{{\bf S}}

\def\IZ{{\bf Z}}

\def\IT{{\bf T}}

\def\tr{{\rm tr \,}}

\title{De Sitter Uplift with Dynamical Susy Breaking}

\author[a,b]{Ander Retolaza}
\author[a]{, Angel Uranga}

\affiliation[a]{Instituto de F\'isica Te\'orica UAM-CSIC \\
C/ Nicol\'as Cabrera 13-15, Campus de Cantoblanco,  28049 Madrid, Spain}
\affiliation[b]{ Departamento de F\'isica Te\'orica, Universidad Aut\'onoma de Madrid,\\ Campus de Cantoblanco, 28049 Madrid, Spain }
\emailAdd{ander.retolaza@uam.es, angel.uranga@uam.es}

\abstract{We propose the use of D-brane realizations of Dynamical Supersymmetry Breaking (DSB) gauge sectors as sources of uplift in compactifications with moduli stabilization onto de Sitter vacua. This construction is fairly different from the introduction of anti D-branes, yet allows for tunably small contributions to the vacuum energy via their embedding into warped throats. The idea is explicitly exemplified by the embedding of the 1-family $SU(5)$ DSB model in a local warped throat with fluxes, which we discuss in detail in terms of orientifolds of dimer diagrams. 
}

\preprint{
\begin{flushright}IFT-UAM/CSIC-15-138 \\ FTUAM-16-1\end{flushright} \vspace{-0.9cm}
}

\begin{document}

\maketitle


\section{Introduction and Conclusions}

The construction of de Sitter vacua in string theory is a most fertile industry in the field. The prototypical approach \cite{Kachru:2003aw} is to consider string compactifications with full moduli stabilization (by fluxes \cite{Dasgupta:1999ss,Giddings:2001yu,Becker:2001pm}, non-perturbative effects, and possibly $\alpha'$ corrections \cite{Balasubramanian:2005zx}), and add some additional sector introducing tunably small contributions to the vacuum energy, to motivate the existence of vacua with parametrically small cosmological constant. This additional sector, typically producing the `uplift' of an originally anti de Sitter stabilized vacuum to a de Sitter one, was introduced in \cite{Kachru:2003aw} as a sector of anti D3-branes at the bottom of a warped throat, whose redshift suppression underlies the claimed tunability. Since then, several alternatives have been proposed \cite{Burgess:2003ic,Bergshoeff:2015jxa,Kallosh:2015nia,Kallosh:2015tea,Cicoli:2015ylx} (see \cite{Saltman:2004jh,Westphal:2006tn,Cicoli:2012fh,Louis:2012nb,Rummel:2014raa,Braun:2015pza} for alternative ideas not involving uplifting sectors of this kind).

A general difficulty in these approaches is that the uplift is not well described within effective field theory (see however \cite{Bergshoeff:2015jxa,Kallosh:2015nia,Kallosh:2015tea} for some improvement in this respect). In this paper we propose a class of models where the uplifting mechanism is built-in in the effective field theory, since it arises from the spontaneous susy breaking dynamics of strongly coupled gauge theory sectors on D3-branes (DSB sectors). Therefore,  our setup is free from   potential problems  arising from limitations of an effective field theory description. The DSB sector is  located at the bottom of warped throats in order to achieve the tunability of the susy breaking contribution to the vacuum energy.\footnote{Another mechanism to uplift the cosmological constant via DSB in the Large Volume Scenario was proposed in \cite{Cicoli:2012fh}. } Although this idea is very general, we particularize our discussion to a concrete explicit model, in which the DSB sector is the susy breaking 1-family $SU(5)$ theory in \cite{Affleck:1983vc}, whose D-brane embedding appeared in \cite{Franco:2007ii}. This particular model has the advantage of having a unique trivial vacuum that becomes non-supersymmetric due to quantum effects \cite{Affleck:1983vc}. This ensures that our setup lies on the true vacuum of the theory, as opposed to other proposals to uplift the cosmological constant \cite{Kachru:2002gs,Saltman:2004jh}.  We describe the D-brane embedding and the corresponding throat geometry, and provide an explicit holographic dual in terms of a duality cascade generalizing the Klebanov-Strassler RG flow \cite{Klebanov:2000hb}. Although the DSB model is of the `non-calculable' kind, similar constructions can be carried out for other DSB D-brane gauge theories in the market \cite{Berenstein:2005xa,Franco:2005zu,Bertolini:2005di,Franco:2006es,Argurio:2006ny, Argurio:2007qk}.

One interesting advantage of the proposal is that the holographic description of the strong gauge dynamics underlying the DSB and thus the uplift lies, by construction, beyond  the regime of validity of 10d supergravity. Therefore this uplift mechanism possibly circumvents problems associated to the description of antibranes and their backreaction in supergravity (for a recent discussion, see \cite{Bena:2014jaa,Bena:2015kia} and references therein).

The paper is organized as follows. In section \ref{sec:dimers-orientifolds} we introduce the description of gauge theory sectors on D3-branes at toric singularities, and orientifolds thereof, using dimer diagrams (section \ref{sec:dimer-intro}), and use it to describe our DSB sector example (section \ref{DSB}) based on an orientifolded $\IC^3/\IZ_6$ singularity. In section \ref{sec:host} we embed this DSB sector as the IR remnant after partial confinement of a host theory, corresponding to D3-branes at a worse singularity which admits a complex deformation to  $\IC^3/\IZ_6$. The general discussion of such complex deformations is carried out in section \ref{sec:hosting-complex}, and applied to the DSB theory in section \ref{sec:hosting-dsb}; the gauge theory analysis of the partial confinement is discussed in section \ref{sec:confine-to-dsb}. Section \ref{sec:warping-down} describes the embedding of the host theory in a warped throat, in terms of the hologarphic dual RG duality cascade. Section \ref{sec:hw-cascade} describes a simpler cascade in terms of a Hanany-Witten brane configuration, and the actual example of interest is detailed in section \ref{sec:cascade}. Finally, appendix \ref{sec:technology} spells out some details on some Seiberg dualities arising in the main text.

\bigskip

\section{Dynamical Susy Breaking from D-branes at singularities} \label{sec:dimers-orientifolds}

\bigskip

\subsection{Dimers and their Orientifolds}
\label{sec:dimer-intro}

\begin{figure}[htb]
\begin{center}
\includegraphics[scale=.23]{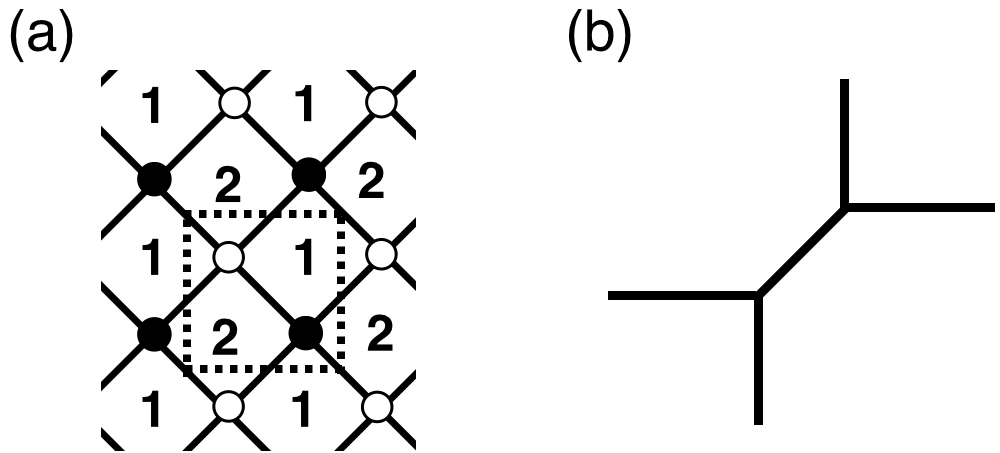} 
\caption{\small (a) Dimer diagram for the theory of D-branes at a conifold. The dashed line is the unit cell in the periodic array. (b)  Web diagram of the conifold. We have displayed it with a finite size $\IS^2$ (middle segment) for clarity; the actual singularity arises when this $\IS^2$ is blown-down.}
\label{fig:conifold-dimer2}
\end{center}
\end{figure}

The gauge theories for D3-branes at toric CY threefold singularities are nicely encoded in a combinatorial graph known as dimer diagram \cite{Franco:2005rj,Franco:2005sm} (see also \cite{Kennaway:2007tq,Yamazaki:2008bt} and references therein). They are (bipartite) graph tilings of $\IT^2$, whose faces correspond to gauge factors, edges represent chiral multiplets in bi-fundamental representations of the neighbouring gauge factors (oriented e.g. clockwise around black nodes, and counterclockwise around white nodes), and nodes representing superpotential couplings (with sign determined by the node color). The diagram for the conifold is shown in figure \ref{fig:conifold-dimer2}(a).
The corresponding gauge theory \cite{Klebanov:1998hh} has gauge group $U(n_1)\times U(n_2)$, bi-fundamental chiral multiplets in two copies of the representation $(\fund_1,\antifund_2)+(\antifund_1,\fund_2)$, denoted by $A_i$, $B_i$, $i=1,2$, and a superpotential $W=\epsilon_{ik}\epsilon_{jl} A_iB_jA_kB_l$.
 
  The geometric information about the CY singularity at which the D-branes are sitting is easily reconstructed from simple combinatorial tools in the dimer, whose discussion we skip, directing the interested reader to the references. Let us simply mention that the geometries are encoded in web diagrams, which specify the fibration structure of the corresponding toric geometry. The web diagram for the conifold is shown in figure \ref{fig:conifold-dimer2}(b).

\begin{figure}[htb]
\begin{center}
\includegraphics[scale=.17]{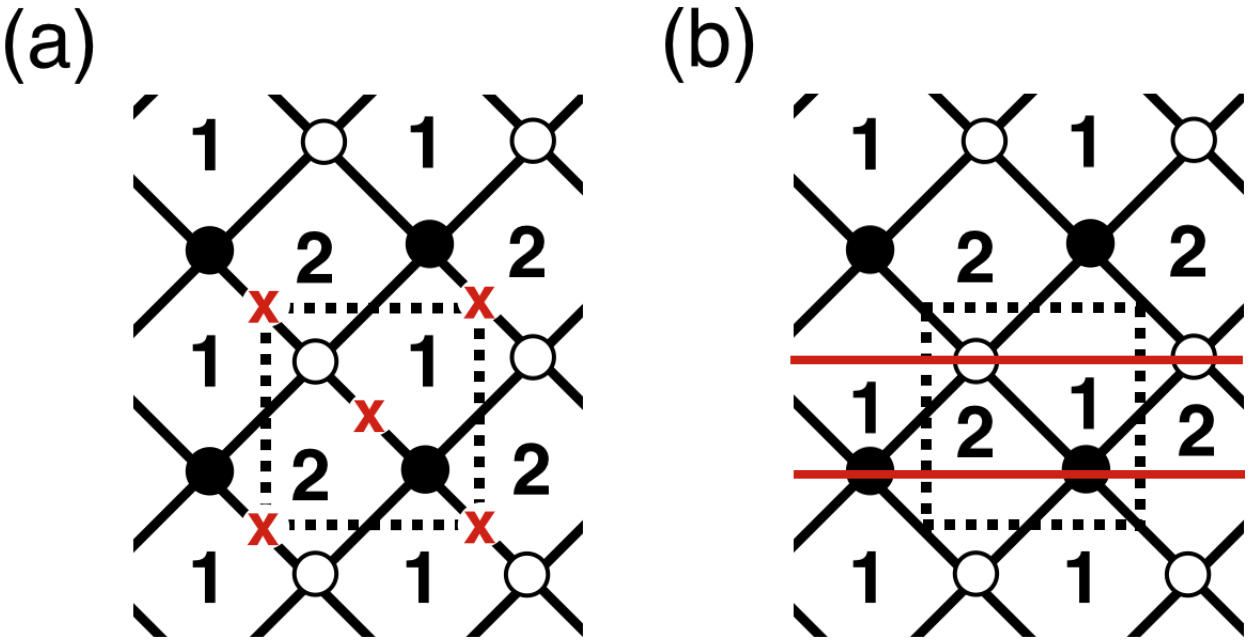} 
\caption{\small Dimer diagram for orientifolds of the conifold with fixed points (a) or fixed lines (b).}
\label{fig:coni-orientifold}
\end{center}
\end{figure}

The combinatorial power of dimer diagrams can be generalized to describe the gauge theories on D-branes at orientifolds of toric singularities. The general description was provided in \cite{Franco:2007ii}, and corresponds to modding out the dimer diagram by a $\IZ_2$ involution. There are two kinds of orientifold quotients, classified by their fixed sets being lines or points. Two such orientifolds of the conifold theory are shown in figure \ref{fig:coni-orientifold}. It is easy to construct other examples, see later and \cite{Franco:2007ii}.

In the following we mainly focus on models with orientifold fixed points in the dimer. For this class, the rules are as follows (see \cite{Franco:2007ii} for detailed derivations). Each orientifold point carries a $\pm$ sign, with the constraint that the number of orientifold planes with the same sign is even (resp. odd) for dimers with number of nodes given by $4k$ (resp. $4k+2$). Orientifold points with charge $+$ (resp $-$) in the middle of a dimer face project down the corresponding gauge factor to $SO(n_a)$ (resp $USp(n_a)$ ). Orientifold points with charge $+$ (resp. $-$) in the middle of a dimer edge project down the corresponding bifundamental onto the two-index symmetic (resp. antisymmetric) representation. Finally, faces and edges not mapped to themselves by the orientifold, combine with their images and descend to $U(n_a)$ gauge factors and bi-fundamental matter multiplets in the orientifold theory. 

\bigskip

\subsection{A Dynamical Susy Breaking Model} 
\label{DSB}

The introduction of orientifolded dimers allows the embedding of D-brane systems realizing Dynamical Susy Breaking at low energies. We will focus on the realization in \cite{Franco:2007ii} of the DSB model described in \cite{Affleck:1983vc} \footnote{For other D-brane systems related to DSB, see also \cite{Berenstein:2005xa,Franco:2005zu,Bertolini:2005di,Franco:2006es,Argurio:2006ny,Argurio:2007qk}.}.

\begin{figure}[ht]
\begin{center}
\includegraphics[scale=.22]{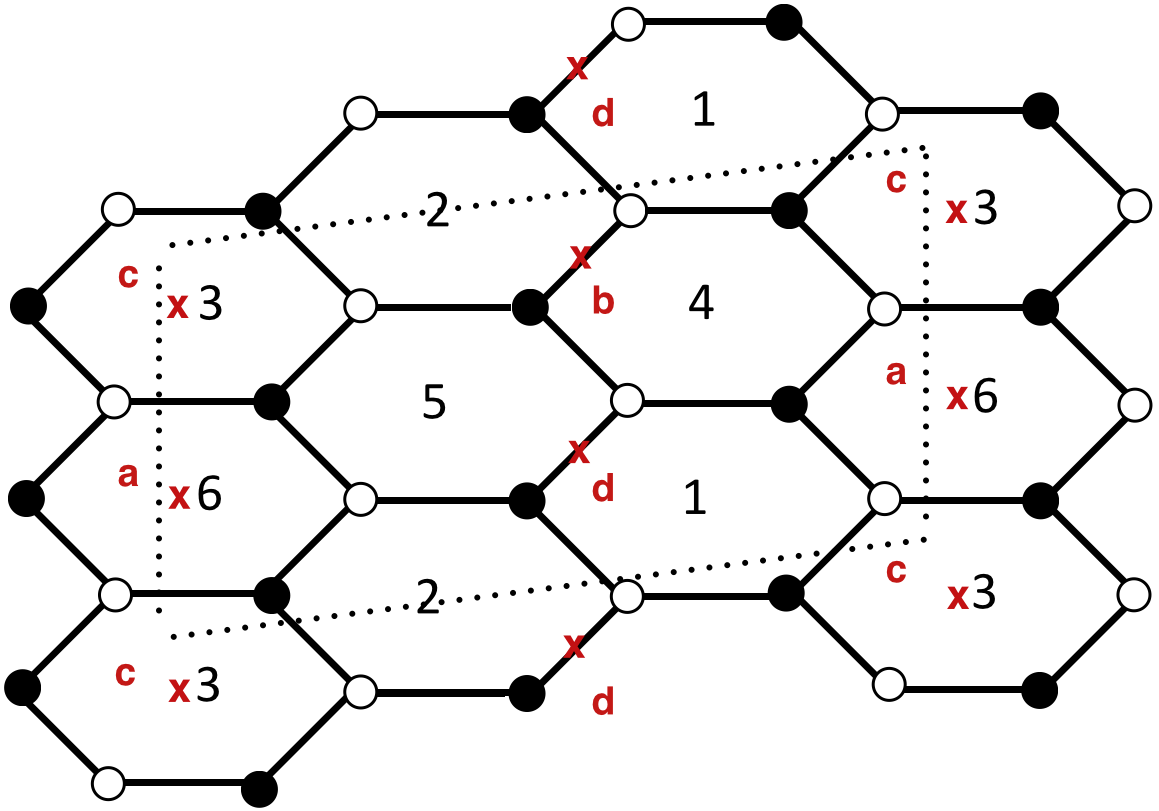}
\caption{\small  Dimer  of $\IC_3/\IZ'_6$ with orientifold points. The labels $a,b,c,d$ are used in the main text to list the orientifold point sign assignment.}
\label{fig:dimer-c3z6-IR}
\end{center}
\end{figure}

The model  is based on the $\IC^3/\IZ_6'$ geometry, where the $\IZ_6'$ is generated by
\beqa
z_i \to e^{2i\pi v_i} z_i,
\eeqa
with $v=(1,2,-3)/6$.  
The orientifolded dimer diagram is shown in  figure \ref{fig:dimer-c3z6-IR}. With a suitable choice of orientifold point signs   (a,b,c,d)=(+ + - - ), the gauge theory for general ranks is
\beqa
&SO(n_6)\times U(n_1)\times U(n_2)\times Sp(n_3) & \nonumber \\
& (\fund_6,\antifund_1) + (\fund_1,\antifund_2) + (\fund_2,\antifund_3)+& \nonumber \\
& + (\fund_6,\antifund_2) + (\fund_1,\fund_3) + \asymm_2+\antisymm_1 +& \nonumber \\
&+  [\, (\fund_6,\fund_3) + (\fund_1,\fund_2) + (\antifund_1,\antifund_2) \,]&.
\eeqa 
As discussed in \cite{Franco:2007ii}, the orientifold group is $(1+\theta+\ldots +\theta^5) (1+\Omega \alpha (-1)^{F_L})$, where $\alpha$ acts as
\beqa
(z_1,z_2,z_3)\to (e^{2i\pi/12}, e^{4i\pi /12}, e^{-6i\pi /12}).
\eeqa
The RR tadpole cancellation condition reads
\beqa
-n_6 + n_2+n_3-n_1-4=0.
\eeqa
It can be solved by e.g. $n_1=n_3=0$, $n_6=N$, $n_2=N+4$. The gauge theory then becomes $SO(N)\times U(N+4)$ with 
matter $(\fund,\antifund)+(1,\asymm)$ and vanishing superpotential. The rank assignment
\beqa
n_1=n_3=n_5=0 \quad ; \quad n_6=1 \quad ; \quad n_2=n_4=5 
\eeqa
produces a theory with gauge group $SO(1)\times U(5)$ and matter content $(\fund , \antifund ) + ( 1, \asymm )$. The $U(5)$ gauge factor is actually $SU(5)$ since the $U(1)$ gauge factor is anomalous and becomes massive after coupling to a RR field. Hence we get a $SU(5)$ theory 
with chiral multiplets in the $10+{\overline 5}$ and no superpotential. This theory has been argued to show dynamical supersymmetry breaking \cite{Affleck:1983mk,Affleck:1983vc}.

The argument for susy breaking is indirect, but compelling. Basically, the theory has a unique vacuum (i.e. trivial moduli space), and the anomaly matching conditions are so strong that cannot be satisfied by any reasonable supersymmetric spectrum. Therefore, susy breaking appears as the simplest option for the low-energy dynamics. In the following we follow the common belief that this in indeed the case. The key idea in the present paper is to bring this DSB system to a form usable as uplifting mechanism in attempts to realize de Sitter vacua in string theory, a la KKLT.

The indirect description of susy breaking implies that the IR theory is strongly coupled, and therefore non-calculable. This is hardly a drawback for uplifting applications, which only require information about the scaling the vacuum energy contribution with parameters. It therefore suffices to introduce a dynamical scale $\Lambda_{\rm DSB}$ to describe the scale of susy breaking strong dynamics, which also controls the vacuum energy of the susy breaking minimum in the rigid limit. The main extra input one needs is the embedding of this DSB sector in a warped throat to redshift this scale, and render the vacuum energy tunable.

\bigskip

\section{Embedding DSB D-branes in a cascading host}
\label{sec:host}

In order to use this kind of DSB sector to uplift with a tunable cosmological constant, it is crucial that the DSB scale is tunable and hierarchically suppressed with respect to the bulk scales. This can be achieved, as in the original anti-D3-brane setup in \cite{Kachru:2003aw}, by locating the DSB brane sector at the bottom of a warped throat.

In field theory terms, one needs a field theory admitting a cascade of Seiberg dualities whose infrared end is the DSB sector of interest. This seems problematic, because the quiver realizing the DSB theory is non-cascading by itself. However, there is a way out recently exploited in \cite{Franco:2015kfa} (see \cite{Cascales:2005rj,Franco:2008jc} for earlier implementations of this idea), which proceeds in two steps. The first, studied in this section, is to realize the DSB theory as the IR description of a suitable larger theory (which we dub the `host'), eventually admitting a duality cascade. The second, studied in the next section, is the description of the duality cascade for the host theory.

 \subsection{Hosting by complex deformation}
 \label{sec:hosting-complex}

The basic tool to embed the DSB into a host theory is through complex deformations of the underlying geometry. Namely, the host theory is realized as a D-brane system at a singularity, which admits a partial smoothing by complex deformation (i.e. growing one or several 3-cycles) to the DSB configuration. As will be clear later on, this complex deformation is actually triggered by the presence of suitable 3-form fluxes.
The field theory counterpart is that the host gauge theory experiences partial confinement, and the resulting light degrees of freedom correspond to the hosted DSB theory. In this section we describe the key ideas of the description of complex deformations, and the corresponding confinement in the D-brane gauge theories, referring to \cite{Franco:2005fd,GarciaEtxebarria:2006aq} for further discussions.

\begin{figure}[htb]
\begin{center}
\includegraphics[scale=.5]{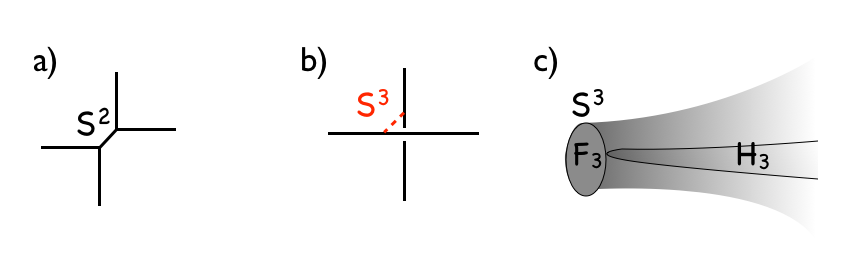} 
\caption{\small (a) The web diagram for the resolved conifold; the finite segment corresponds to the 2-sphere. (b) The splitting of the diagram into sub-webs in equilibrium describes the complex deformation; the dashed segment measuring the sub-web separation represents the 3-sphere. (c) The 3-sphere in the complex deformation can support fluxes which lead to the KS warped throat.}
\label{fig:conifold}
\end{center}
\end{figure}

As explained in \cite{Franco:2005fd}, the description of complex deformations for toric singularities is very easy. It corresponds to a splitting of the web diagram into sub-webs in equilibrium. The resulting sub-webs encode the toric description of the singular points possibly remaining after the smoothing by complex deformation.
Let us review the prototypical case of the conifold, whose web diagram is shown in figure \ref{fig:conifold}(a). Its splitting into sub-webs in equilibrium (in this case, just lines) is shown in figure \ref{fig:conifold}(b).

In this case, the complex deformation results in a smooth space. In our field theory terms, the conifold theory acts as a host for the $\NN=4$ SYM theory of D3-branes in flat space. The analysis is as follows: the complex deformation in the geometry relates (as will be discussed later on, but is familiar from \cite{Klebanov:2000hb}) to the strong dynamics of the conifold field theory with ranks corresponding to $SU(2M)\times SU(M)$. The first factor has $N_f=N_c$ and has a quantum deformed moduli space, which is eventually parametrized by mesons $M_{ij}=\tr(A_iB_j)$, subject to the constraint $\det M= \Lambda^4$, where $\Lambda$ is the dynamical scale. The gauge group is broken to just $SU(M)$ and the light matter fields correspond to three unconstrained adjoint chiral multiplets with cubic superpotential, corresponding to $\NN=4$ SYM.

In order to construct more general throats, we must consider more general toric singularities, which are easily engineered using the web diagrams. The quiver gauge theories associated to D-branes at such singularities can systematically studied using dimer diagram techniques. However, not all such geometries admit complex deformations, and therefore not all can be used to define warped throats with smooth infrared ends. The criterion for the existence of complex deformations, and their dual field theory interpretation, were described in \cite{Franco:2007ii}. The result is that complex deformations correspond to splitting the web diagram into sub-webs in equilibrium. The field theory dual is described in terms of confinement on a set of gauge factors associated to certain fractional branes, and was systematically studied in terms of dimer diagrams in \cite{Franco:2007ii,GarciaEtxebarria:2006aq}. 

The geometries admitting complex deformations can be used to build throats, which are supported by (2,1) 3-form fluxes, with RR fluxes on the 3-cycles at the bottom of the throat, and NSNS fluxes on their dual 3-cycles. The field theory duals of these throats correspond to duality cascades triggered by the fractional branes dual to the RR 3-form fluxes, which lead to a reduction of the effective number of D3-branes as one runs to the infrared, and which ends in a process of confinement at a dynamical scale dual to the size of the infrared 3-cycles. The explicit construction of metrics for these throats requires information about metrics for the corresponding (deformed) cones (see \cite{Franco:2004jz} for partial results in some examples). However, the main properties of the throats, like the existence of (2,1) fluxes, and the scaling of the warp factor with the flux quanta, can be established even with no information about the underlying metric. These properties are easily encoded in the existence of a supersymmetric RG flow representing a duality cascade.

\subsection{Hosting the DSB sector}
\label{sec:hosting-dsb}

As an application of the previous knowledge, we will take a dimer with orientifold points from \cite{Franco:2007ii} that leads to dynamical supersymmetry breaking and UV complete it by placing it as the IR endpoint of a more complicated gauge theory where several groups confine.  

\begin{figure}[ht]
\begin{center}
\includegraphics[scale=0.3]{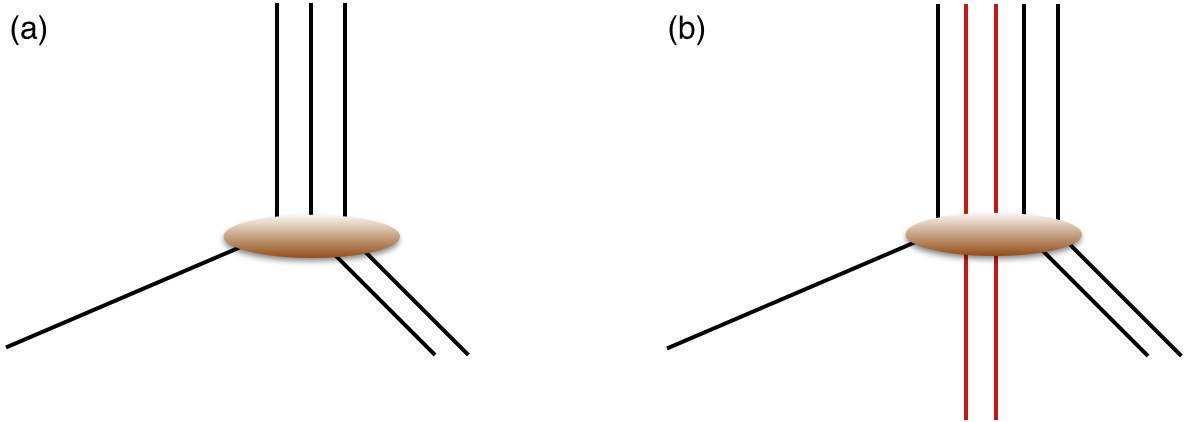}
\caption{\small  (a) External legs of the web diagram of the $\IC_3/\IZ'_6$ orbifold whose dimer is given by figure \ref{fig:dimer-c3z6-IR}. (b) The addition of one line and its orientifold image provide the web diagram for a host configuration, which by construction can be deformed back to $\IC_3/\IZ'_6$.}    
 \label{fig:web-c3z6}
\end{center}
\end{figure}

The web diagram of this orbifold is given by   figure \ref{fig:web-c3z6}(a). To host this theory, we introduce two additional  parallel lines (eventually related by the orientifold projection) in the same direction as one of the external legs of the web diagram to obtain the host geometry, as shown in  figure \ref{fig:web-c3z6}(b). By construction it admits a complex deformation to the $\IC^3/\IZ_6'$ singularity, by removal of the two lines.

\begin{figure}[ht]
\begin{center}
\includegraphics[scale=.3]{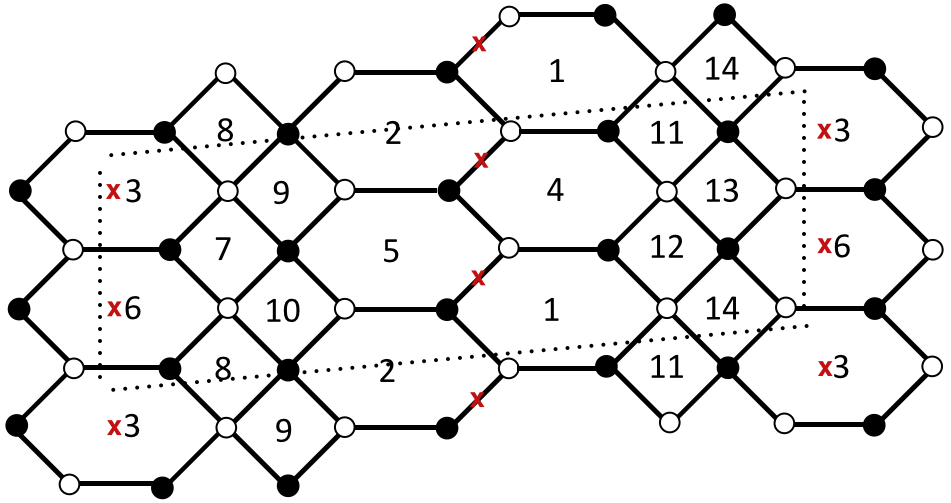}
\caption{\small  Dimer of the host theory for the DSB sector of D-branes at the $\IC_3/\IZ'_6$ orientifold.}
\label{fig:dimer-c3z6-UV}
\end{center}
\end{figure}

The dimer corresponding to the web in  figure \ref{fig:web-c3z6}(b) is shown in  figure \ref{fig:dimer-c3z6-UV}. The two lines added in the web diagram correspond to the new columns of rhombi that are each other's orientifold image. 

\subsection{Confinement onto the DSB theory}
\label{sec:confine-to-dsb}

\begin{figure}[ht]
\begin{center}
\includegraphics[scale=.3]{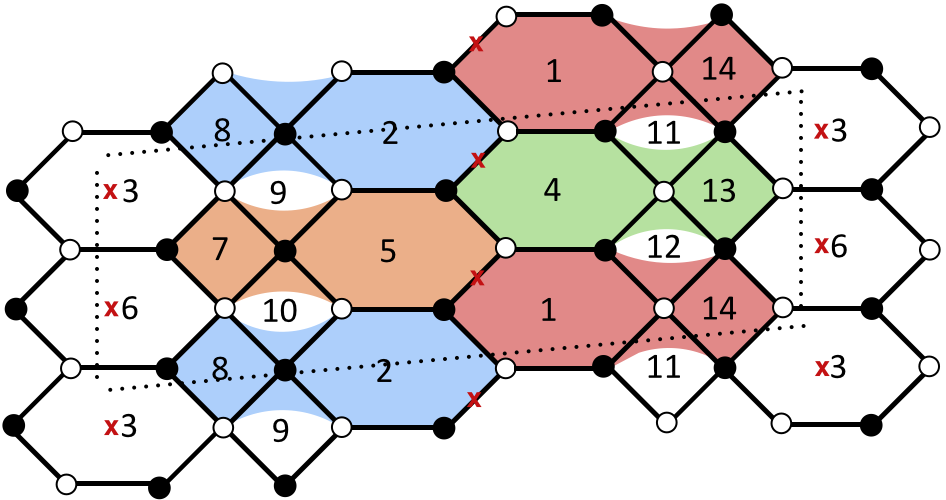}
\caption{\small  Dimer description of partial confinement in the host theory, leaving the DSB dimer theory at low energies. Confinement of the nodes 9, 10, 11 and 12 triggers non-zero vevs for suitable mesons, which  higgs certain gauge factors to the diagonal combinations, depicted as the recombination of dimer faces shaded with same color. The resulting theory is easily seen to agree with figure \ref{fig:dimer-c3z6-UV}.}
\label{fig:confine}
\end{center}
\end{figure}

For suitable rank choices, the host theory experiences confinement of some of its gauge factors, after which the effective field theory describing the light degrees of freedom is precisely the DSB theory. This confinement pattern can be easily studied using the dimer diagram, as shown in figure \ref{fig:confine}. Specifically, the nodes 9, 10, 11 and 12 confine, and force some of their mesons to acquire non-zero vevs. This triggers breaking of their respective flavour symmetry factors (corresponding to neighbouring faces in the dimer) to the diagonal combination. The resulting theory si described by the dimer with recombined faces, which is easily recognized as that in figure \ref{fig:dimer-c3z6-UV}. The fact that this confinement corresponds to the removal of lines in the web diagram, c.f. figure \ref{fig:web-c3z6} can be easily shown using the techniques in \cite{GarciaEtxebarria:2006aq}.

\medskip

Let us describe in some more detail our case of interest, in which the host theory confines onto the DSB theory. The rank choices of the host theory that reproduce the correct rank assignments of the DSB theory in section \ref{DSB} are
\begin{align}
n_1=n_3=n_5=n_7=n_{14}=0 \quad  ; \quad  n_{10}=n_{12}=  6+M \nn \\
n_2=n_4=n_8=n_{13}=5 \quad ; \quad n_{11}=n_9=5+M \quad ; \quad n_6=1  \ , \label{ranks-UV_DSB}
\end{align}
with $M\gg 1$ to obtain a large deformation.

The superpotential at this point can easily be read from the dimer after taking into account that many faces are empty:
\beq
W= X_{9,8} X_{8,10} X_{10,2} X_{2,9} \ .
\eeq

At this point gauge groups 9 and 10 and their orientifold images 11 and 12 have $N_c>N_f$ and thus confine,  developing   ADS superpotentials. The dynamics of both groups is described in terms of their mesons. Looking at the dimer one would expect the mesons to be
\begin{align}
{\cal M }= \ & \left( \begin{matrix}
    M_{82} & M_{87}  \\
    M_{52} & M_{57}  
\end{matrix} \right)=\left( \begin{matrix}
    X_{8,10}X_{10,2} & X_{8,10}X_{10,7}  \\
 X_{5,10}X_{10,2} & X_{5,10}X_{10,7} 
\end{matrix} \right) \\
{\cal N }= \ & \left( \begin{matrix}
    N_{28} & N_{25}  \\
    N_{78} & N_{75}  
\end{matrix} \right)=\left( \begin{matrix}
    X_{29}X_{98} & X_{29}X_{95}  \\
 X_{79}X_{98} & X_{79}X_{95} 
\end{matrix} \right) \, 
\end{align}
but since several nodes have rank zero, there are no mesons charged under these gauge groups. This leaves really one meson for each confining gauge group:
\begin{equation}
{\cal M } = M_{82}= X_{8,10}X_{10,2} \quad ; \quad {\cal N } = N_{28}= X_{29}X_{98} \ .
\end{equation}
The superpotential, taking into account the strong dynamics of these groups, becomes
\beq
W=  {\cal M} {\cal N} + M\left( \det {\cal N} \right)^{-1/M}  + (M+1)\left( \det {\cal M} \right)^{-1/(M+1)} \ .
\eeq
Using the F-terms of ${\cal M} $ and ${\cal N}$ it is easy to see that both get a vev. This breaks gauge groups $SU(n_8)\times SU(n_2)$ and $SU(n_7)\times SU(n_5)$ to their diagonals, and the same happens with their orientifold images. The resulting theory agrees with the recombined dimer which describes the DSB theory.

\section{Warped down DSB using the host cascade}  
\label{sec:warping-down}
  
  The host theory described above admits a duality cascade, which provides the holographic dual of the embedding of the DSB sector in a warped throat. In this section we describe this cascade.
  
Initially,   we work for simplicity in the UV theory, with large numbers of branes, so we ignore the ${\cal O}(1)$ effects due to orientifolds, etc. This simplifies the analysis and displays the cascade structure more nicely. Later on we consider the dualities near the IR of the cascade, where orientifold contributions are non-negligible.
  
 The theory belongs to a general class, closely related to class ${\cal S}_k$ theories in \cite{Gaiotto:2015usa,Franco:2015jna,Hanany:2015pfa}. The cascade involves dualities involving whole columns of hexagons/rhombi, already explored in these references. The structure is inherited from a parent theory, which admits a description in terms of HW configurations. We study this first.
 
\subsection{Parent HW cascade} 
\label{sec:hw-cascade}

\subsubsection{The HW configuration}

\begin{figure}[ht]
\begin{center}
\includegraphics[scale=.2]{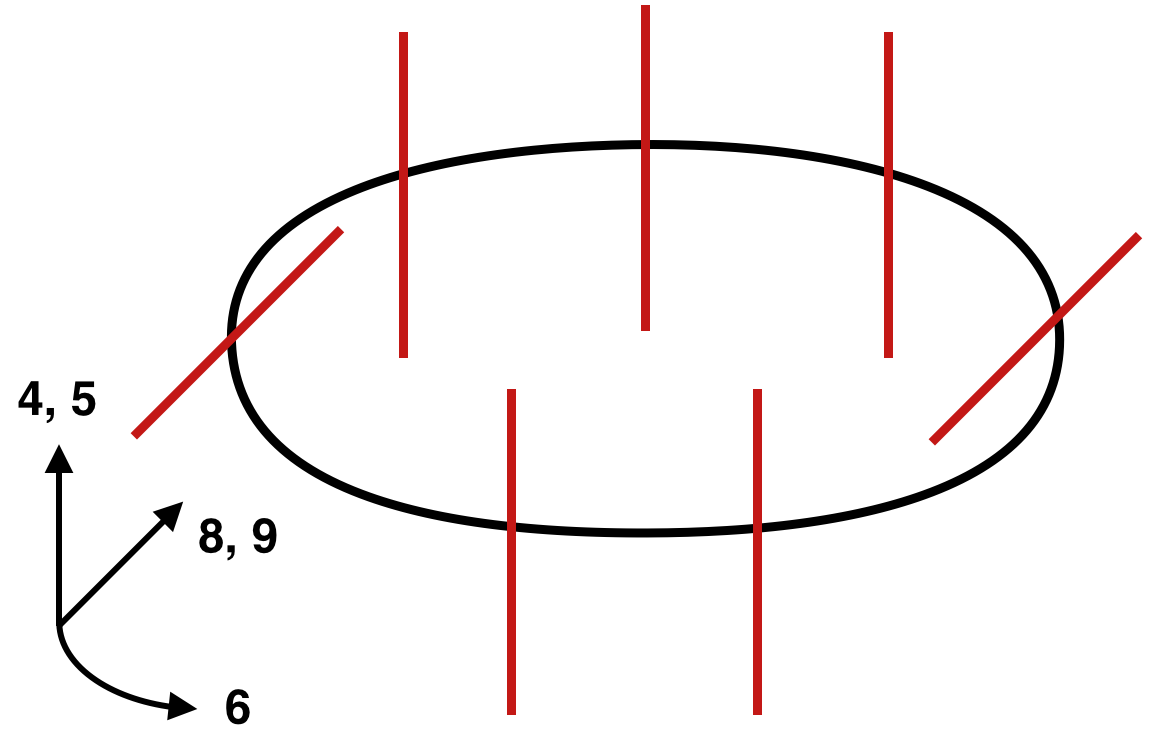}
\caption{\small  Configuration of NS-, NS'- and D4-branes.}
\label{fig:HW-c3z6-UV}
\end{center}
\end{figure}

We consider configuration of NS-branes along 012345, NS'-branes along 012389 and D4-branes along 0123 and suspended in 6 along intervals between the 5-branes. The configuration of interest is shown in figure \ref{fig:HW-c3z6-UV}. By T-duality along the $\IS^1$ direction $x^6$, one recovers a system of D3-branes at a singularity $xy=z^2w^5$, see \cite{Uranga:1998vf}.

In this picture the cascade is nicely described because Seiberg dualities are brane crossings \cite{Elitzur:1997fh}. Thus, the cascade corresponds to moving the NS'-branes around the circle, crossing the NS-branes. Since we are interested in orientifolded theories, the two NS'-branes will move in opposite directions in a $\IZ_2$ symmetric way. The resulting picture is related to the cover space description of the eventual orientifold cascade.
\begin{figure}[ht]
\begin{center}
\includegraphics[scale=.47]{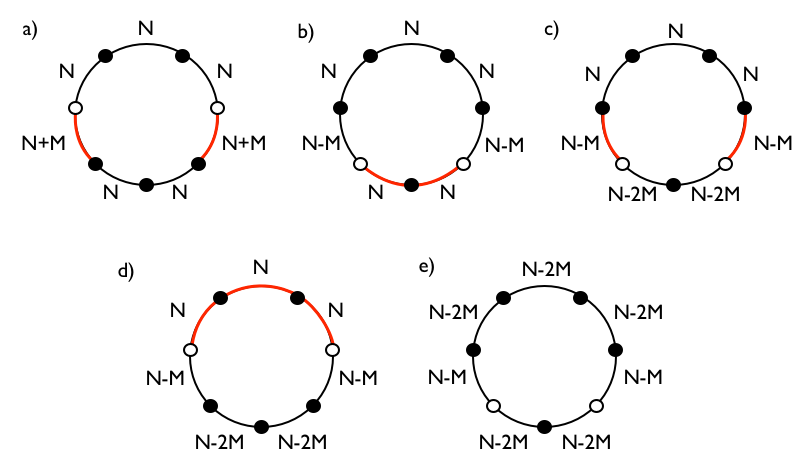}
\caption{\small Structure of the basic period of the HW cascade. Black and white circles denote the two kinds of 5-branes. The gauge factors dualized in each step are shown as red intervals. After four steps, we recover the original configuration with a reduced number $N'=N-2M$ of regular D-branes, and the same number $M$ of fractional branes.}
\label{fig:hw-cascade}
\end{center}
\end{figure}

The choice of ranks and the basic period of the cascade are shown in figure \ref{fig:hw-cascade}.  At the level of the brane configuration, the changes in the numbers of D4-branes are determined by conservation of the linking number \cite{Hanany:1996ie} for NS- and NS'-branes. Recall that the linking number for a 5-brane is given (in this simple configurations) by the difference in the numbers of D4-branes ending on it from both sides. This number corresponds to a vortex number from the viewpoint of the 5-brane worldvolume theory, and is conserved in any local deformation, including brane crossings.

From the viewpoint of the 4d field theory, the brane motions sometimes correspond to some tricky Seiberg dualities, which we discuss in appendix \ref{sec:technology}.

\subsection{The cascade in the DSB host theory} \label{sec:cascade}

\subsubsection{The far UV}

Let us now describe the duality cascade of the host theory for our DSB sector. We start with the discussion in the UV, where the numbers of regular and fractional branes are large compared with the orientifold plane charge. Hence, the effects of the orientifold are suppressed and can be neglected, save for the fact that the cascade should respect the $\IZ_2$ symmetry.

As we will show, the cascade pattern follows very easily from the cascade of the HW parent theory above. In particular, each node of the HW theory is related to a column of either hexagons or rhombi in the dimer of interest, and the dualization of a HW node by brane crossing is related to the dualization of whole columns in the dimer. This is a particular application of ideas already introduced in \cite{Franco:2015jna}, to which we refer for further details.

Consider the theory described by figure \ref{fig:dimer-c3z6-UV} without the orientifold points. This is directly related to the HW configuration in figure \ref{fig:hw-cascade}(a) with the columns $(1,4)$, $(11,12)$, $(13,14)$, $(3,6)$, $(7,8)$, $(9,10)$ related to the HW intervals according to the following rule: columns of hexagons relate to intervals between parallel 5-branes, while columns of rhombi relate to intervals between orthogonal 5-branes.

According to this relation, we consider the initial ranks at certain point in the UV of the cascade:
\begin{align}
n_1=n_2=n_3=n_4=n_5=n_6=n_7=n_8=n_{13}=n_{14}= \ & N \nn \\
 n_{9}=n_{10}=n_{11}=n_{12}= \ & N+M \label{ranks-before}
\end{align}
with $N\gg M$.

\begin{figure}[ht]
\begin{center}
\includegraphics[scale=.3]{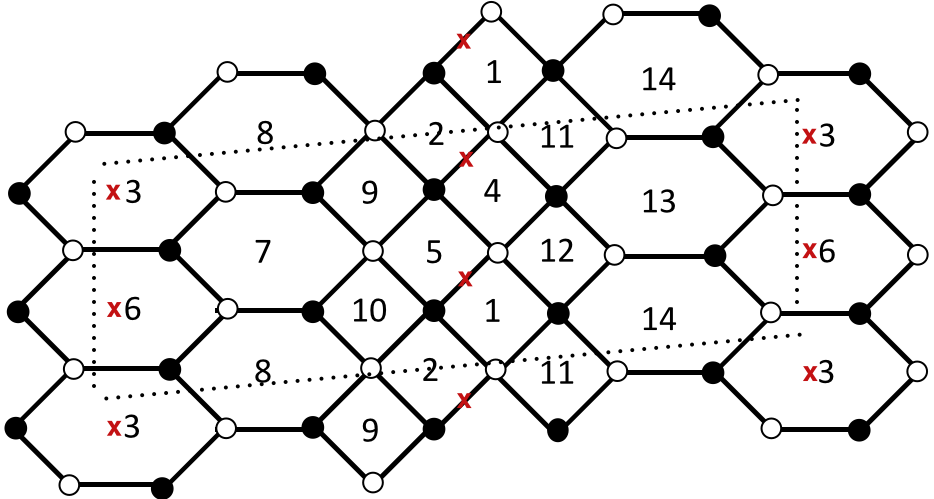}
\caption{\small  Dimer of the DSB host theory after performing Seiberg duality in nodes 9, 10, 11 \& 12.}
\label{fig:dimer-c3z6-UV-2}
\end{center}
\end{figure}

The next step in the cascade is to dualize the nodes 9, 10, 11 \& 12. As discussed in \cite{Franco:2015jna}, such dualizations of columns change the adjacent columns from hexagons to squares (for 1,4 and 2,5) and vice-versa (for 7,8 and 13,14), resulting in the dimer shown in figure \ref{fig:dimer-c3z6-UV-2} (without the orientifold points). 

The ranks after performing the duality are 
\begin{align}
n_1=n_2=n_3=n_4=n_5=n_6=n_7=n_8=n_{13}=n_{14}= \ & N \nn \\
 n_{9}=n_{10}=n_{11}=n_{12}= \ & N-M
\end{align}
in agreement with the result in the parent HW configuration figure \ref{fig:hw-cascade}(b). We skip the detailed discussion of the remaining steps, and simply restate that one obtains a cascade directly inherited from the HW cascade described above.
%

\subsubsection{The near IR regime }

As we approach the IR, the number of regular branes decreases, and eventually becomes not large enough compared with the orientifold charge. Hence, the last steps of the cascade require dealing with the latter. We sketch the corresponding analysis in this section.

Consider the near IR regime theory to be given by the by-now familiar dimer, in figure \ref{fig:dimer-c3z6-UV}, with ranks
\begin{align}
n_1=n_5=6+2M \quad  ;  \quad  n_7=n_{14}=n_3= 6+2M \quad ; \quad  n_{10}=n_{12}=  12 +3M \nn \\   n_2=n_4=11+2M \;\; ;  \;\;
 n_9=n_{11}=11+3M \;\; ;  \;\;  n_8=n_{13}=11 +2M \;\; ;  \;\; n_6=7+2M  \ , 
\end{align}
with $M \gg 1$ . Note that all ranks have increased by $6+2M$ compared to those of (\ref{ranks-UV_DSB}). We start by dualizing nodes 9, 10, 11, 12 following the order in the HW configuration. This dualization has two consequences: on the one hand, it decreases the ranks of these gauge groups according to usual Seiberg duality rules. On the other hand, it changes the shape of the dimer to figure \ref{fig:dimer-c3z6-UV-2}. The ranks after the dualization are:
\begin{align}
n_1=n_5=6+2M \quad  ;  \quad  n_7=n_{14}=n_3= 6+2M \quad ; \quad  n_{10}=n_{12}=  5 +M \nn \\   n_2=n_4=11+2M \;\; ;  \;\;
 n_9=n_{11}=6+M \;\; ;  \;\;  n_8=n_{13}=11 +2M \;\; ;  \;\; n_6=7+2M  \ . 
\end{align}
At this point we can perform double Seiberg duality at nodes 1, 5, 2, 4. Following the ideas in appendix \ref{sec:tech-double},  the dimer after performing the dualization looks the same as the one before the dualization, and the difference between both dual theories lies on the ranks of the dualized gauge groups. The ranks of all these gauge groups after the duality have decreased by an amount $\Delta$, that according to the analysis in appendix \ref{sec:tech-double}  is given by
\beq
\Delta = n_9+n_{10}-n_1-n_2=-6 -2M\ .
\eeq
This change on the ranks of groups 1, 5, 2,  4  leaves the ranks after the double duality
\begin{align}
n_1=n_5=0 \quad  ;  \quad  n_7=n_{14}=n_3= 6+2M \quad ; \quad  n_{10}=n_{12}=  5 +M 
 \nn \\ \! \! \! \! \! \! \! \! \! \! 
n_2=n_4=5 \;\; ;  \;\;   n_9=n_{11}=6+M \;\; ;  \;\;  n_8=n_{13}=11 +2M \;\; ;  \;\; n_6=7+2M  \ . 
\end{align}
Now we dualize faces 9, 10, 11, 12 again to end up with a dimer with the previous shape and ranks 
\begin{align}
& n_1=n_5=0 \quad  ;  \quad  n_7=n_{14}=n_3= 6+2M \quad ; \quad  n_{10}=n_{12}=  6 +M \quad  ;  \quad n_2=n_4=5 \nn \\ 
&\quad n_9=n_{11}=5+M \quad ; \quad  n_8=n_{13}=11 +2M \quad ; \quad n_6=7+2M  \ . 
\end{align}
We now proceed to dualize the hexagon column formed by faces 3 and 6 together with the adjacent rhombi columns with  faces 7, 8 and  their orientifold images 13, 14. Once again, the dimer after the duality has the same shape and decreased ranks in the dualized gauge groups we just mentioned. The change in all these ranks  this time is given by 
\beq
\Delta = n_9+n_{10}-n_7-n_8 = -6-2M \ ,
\eeq
leaving the ranks after the dualization
\begin{align}
n_1=n_3=n_5=n_7=n_{14}=0 \quad  ; \quad  n_{10}=n_{12}=  6+M \nn \\
n_2=n_4=n_8=n_{13}=5 \quad ; \quad n_{11}=n_9=5+M \quad ; \quad n_6=1  \ ,
\end{align}
that, as described in section \ref{sec:confine-to-dsb}, produces upon partial confinement the DSB gauge theory.

\bigskip

\section*{Acknowledgments}

We would like to thank L. Ib\'a\~nez and F. Marchesano for useful discussions. We are also indebted to Sebasti\'an Franco for useful comments while developing previous projects. A. R. and A. U. are partially supported by the grants FPA2012-32828 from the MINECO, the ERC Advanced Grant SPLE under contract ERC-2012-ADG-20120216-320421 and the grant SEV-2012-0249 of the ``Centro de Excelencia Severo Ochoa" Programme. 

\bigskip

\appendix

\newpage

\section{Details on certain Seiberg Dualities} \label{sec:technology}

From figures \ref{fig:dimer-c3z6-UV} and \ref{fig:dimer-c3z6-UV-2} we can observe that the cascade of Seiberg dualities  compatible with the action of the orientifold symmetry involves simultaneous dualization of two neighbouring gauge factors of the dimer, as well as dualizations involving gauge groups represented as hexagons in the dimer, which arise from orbifolding gauge groups with matter in the adjoint representation. As these are somewhat unfamiliar operations, we provide some extra details in this Appendix.

\bigskip

\subsection{Seiberg Duality for neighbouring nodes} \label{sec:tech-double}

Consider the dualization of the DSB host theory corresponding to step (b) to (c) in figure \ref{fig:hw-cascade}, namely simultaneous dualization of faces 1, 2, 4, 5, in figure \ref{fig:dimer-c3z6-UV-2}. The result of the simultaneous dualization does not directly follow from sequential application of dualities; the final result is however easily guessed by simply inheriting it from the corresponding brane crossing in the parent Hanany-Witten configuration. We eventually also discuss the modifications due to the presence of the orientifold points. 

\bigskip

\subsection*{The Hanany-Witten picture}

\begin{figure}[ht]
\begin{center}
\includegraphics[scale=.42]{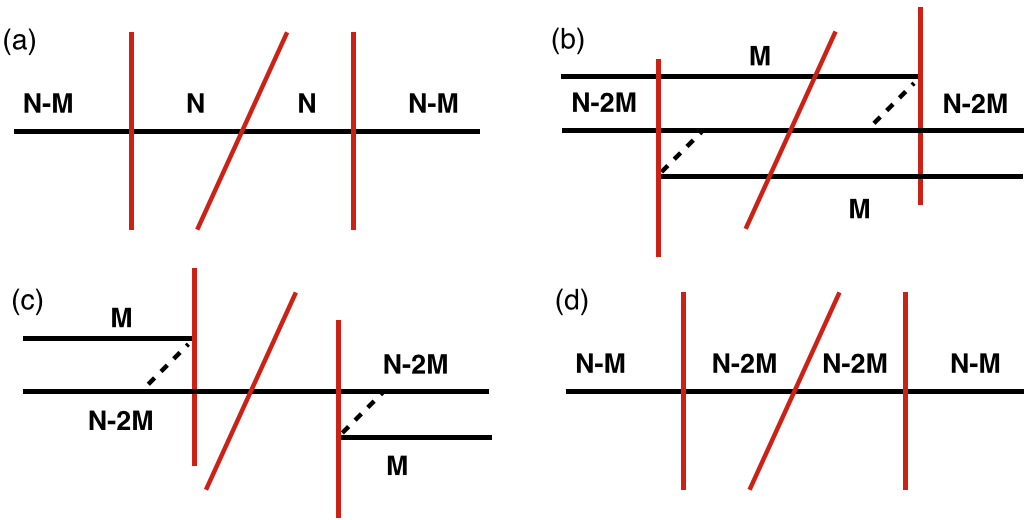}
\caption{\small HW description of the simultaneous Seiberg duality of two neighbouring nodes. In the original theory (a) we move the NS-branes off along with $M$ D4-branes (b). The NS-branes are subsequently moved across the NS'-brane (c), and brought back in line (d).}
\label{tricky-hw1}
\end{center}
\end{figure}

The HW configuration describing our initial gauge theory is shown in figure \ref{tricky-hw1}(a). Vertical and tilted red lines represent NS- and NS'-branes, respectively, while horizontal black lines are D4-branes. 

In the HW picture Seiberg duality is represented as the crossing of a NS5 and a NS5' brane. This is achieved by moving the NS-branes off in say direction 7, together with $M$ semi-infinite D4-branes as shown in figure \ref{tricky-hw1}(b). They are subsequently moved across the NS'-brane, and brought back into position. The resulting theory has the same geometry of NS- and NS'-branes, but the number of D4-branes in the middle segments is decreased. 

The HW configuration easily translates into gauge theory data, allowing to read out the Seiberg dual field theories. The number of D4-branes in the dualized segments decreases by $2M$. This amount arises from the numbers of branes in the dualized  intervals and their neighbours. The general expression relating the ranks  of the two dual theories is
\beq
\tilde{n}_i = \Delta + n_i  \quad ; \quad \Delta= \displaystyle \sum_{\text{adjacent}} n_i  -  \displaystyle \sum_{\text{dualized}} n_j   \ , \label{rank-changes}
\eeq
where $\tilde{n}_i$ stands for the rank of the dualized gauge group after the dualization,  the first sum goes over the ranks of gauge groups not dualized in the process (the ones with  $N-M$ branes in our case) and the second sum goes over the gauge groups we are dualizing (the ones starting with $N$ branes). We see that in our simple example $\Delta=2(N-M)-2N=-2M$, reproducing figure \ref{tricky-hw1}(a) and (d).


\subsubsection*{The dimer picture}

\begin{figure}[h]
\begin{center}
\includegraphics[scale=.22]{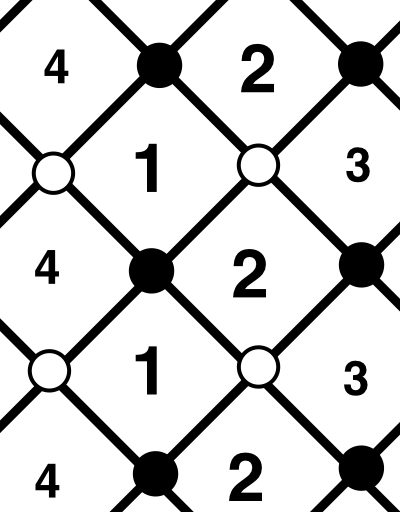}
\caption{\small Dimer diagram view of the HW brane configuration in figure \ref{tricky-hw1}.}
\label{fig:double-seiberg-dimer-simple}
\end{center}
\end{figure}

Translating the previous  HW brane configuration into the dimer picture  is straightforward and will be useful to later derive the rules for more general simultaneous dualization of neighbouring gauge groups on the dimer. In figure \ref{fig:double-seiberg-dimer-simple} we show the  part of a dimer corresponding to the whole HW set-up in figure \ref{tricky-hw1}(a). The two intervals where we performed Seiberg duality translate to columns of rhombi touching each other, that we labeled 1 and 2. We also included the adjacent intervals, that translate to faces with labels 3 and 4 right next to the previous rhombi columns.   The ranks are the same as those in figure \ref{tricky-hw1}(a):
\beq
n_1=n_2=N \quad ; \quad n_3=n_4=N-M \ .
\eeq
%

The fact that figure \ref{tricky-hw1}(a) and figure \ref{tricky-hw1}(d) only differ on the amount of D4-branes implies that the  dimer after the dualization looks the same, but now the ranks are lower for the dualized gauge groups. Using (\ref{rank-changes}) the change in the ranks of the dualized gauge groups is
\beq
\Delta = (n_3+n_4)-(n_1+n_2)=-2M \ ,
\eeq
which leaves the ranks for the dual theory
\beq
n_1=n_2=N-2M \quad ; \quad n_3=n_4=N-M \ .
\eeq

\subsubsection*{Adding orientifold points}

We now deal with the  case of interest  in the main text corresponding to simultaneous dualization of faces 1, 2, 4, 5, in figure \ref{fig:dimer-c3z6-UV-2}. We use a similar approach, with only few additional technicalities 

\begin{figure}[h]
\begin{center}
\includegraphics[scale=.2]{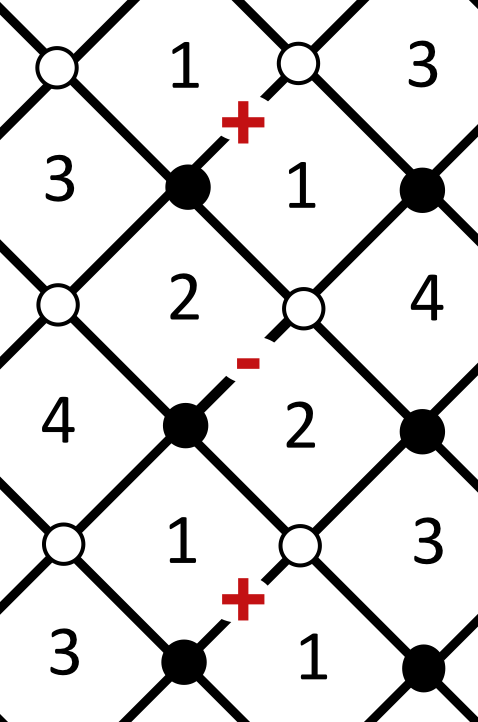}
\caption{\small Orientifolded dimer diagram for a $\IZ_2$ orbifold of a HW configuration. The orientifold point charges are shown in red.}
\label{fig:dimer-double-eiberg-z2-opoints}
\end{center}
\end{figure}

The piece of the dimer of interest is shown in figure \ref{fig:dimer-double-eiberg-z2-opoints}, where we use equal labels for orientifold image faces (and which do not correspond to the labels in the main text, hoping no confusion arises). Note that it does not admit a HW brane configuration description.

A first relevant point is the change in the spectrum and anomalies.  In particular, edges on top of the orientifold points describe chiral multiplets in two-index (anti)symmetric representation of the corresponding gauge group, depending on the orientifold charge. This modifies the anomaly cancellation conditions, and forces gauge factors in the same column to actually have different ranks according to 
\beq
n_4+n_1=n_2+n_3+4 \ . \label{anomalies-z2-opoints}
\eeq 
Upon dualization of 1 and 2, the dimer describing the dual theory looks the same, with the only difference being the ranks of the gauge groups on these columns of rhombi. All gauge groups in the dual theory have ranks related to the initial ones via a unique amount $\Delta$ that we can easily calculate using a generalization of (\ref{rank-changes}). Taking into account the identification between gauge groups due to the orientifold points, the relation between the ranks of both theories is then given by
\beq
\tilde{n}_i=n_i +\Delta  \quad \text{for} \ \  i=1,\ 2  \quad ; \quad \Delta = (n_3+n_4)-(n_1+n_2) \ .
 \eeq

\medskip

\bigskip

\subsection{Seiberg duality involving columns of hexagons} \label{sec:tech-adj}

Consider the dualization of the DSB host theory corresponding to step (d) to (e) in figure \ref{fig:hw-cascade}. In the orbifold dimer diagram, the dualization involves a column of hexagons and the neighbouring columns of rhombi, namely  faces 3, 6, 7, 8, 13, 14 in figure \ref{fig:dimer-c3z6-UV}.
The complete dualization does not follow from naive sequential dualizations of the different columns. Again, we can proceed by using intuitions from brane crossing in the parent HW configuration.

\subsubsection*{The HW picture}

\begin{figure}[ht]
\begin{center}
\includegraphics[scale=.41]{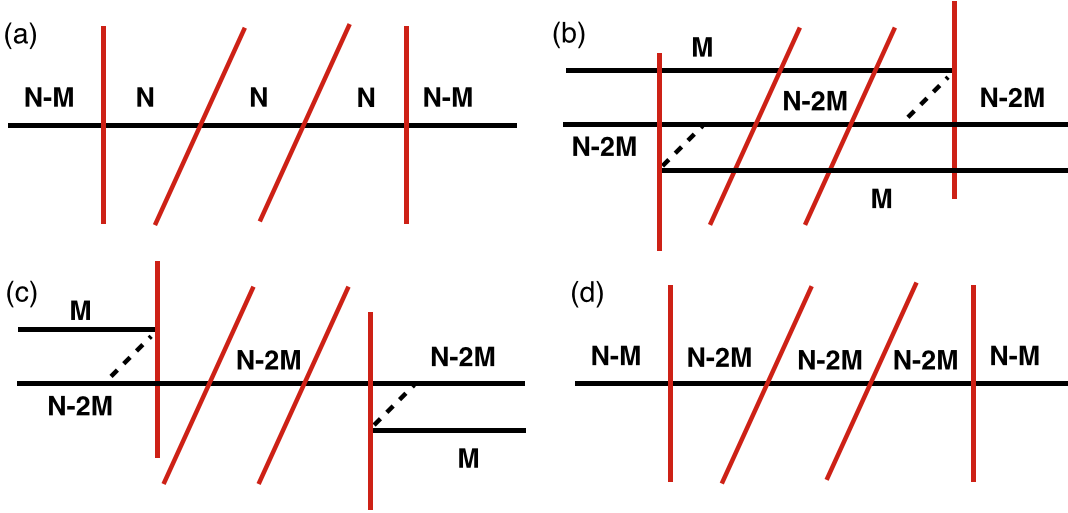}
\caption{\small HW description of the simultaneous Seiberg duality involving an $\NN=2$ sector. In the original theory (a) we move the NS-branes off along with $M$ D4-branes (b). The NS-branes are subsequently moved across the NS'-brane (c), and brought back in line (d).}
\label{tricky-hw2}
\end{center}
\end{figure}

The HW configuration and the brane moves corresponding to the overall Seiberg duality are shown in figure \ref{tricky-hw2}. The hexagon in the dimer corresponds to the gauge group on the stack of D4-branes between the parallel NS'-branes in the middle of figure \ref{tricky-hw2}(a). The dualization process involves not only this gauge group but also the adjacent ones, since in the HW picture it corresponds to moving the NS branes on both sides in opposite directions. In order to describe the dualization, we proceed by first taking the NS-branes to opposite points on direction 7 and then moving them along direction 6 as shown in figure \ref{tricky-hw2}(b) and  \ref{tricky-hw2}(c). Finally, we put the NS-branes back at $x^7=0$  to end up with the configuration in figure \ref{tricky-hw2}(d), that looks very similar to the initial one, the only difference being the amount of D4-branes in all the intervals between both NS-branes. 

Once again, the two brane configurations only differ on the amounts of D4-branes on the intervals we dualized, and therefore the dual gauge theory  differs only on the rank of the three dualized gauge groups. It is  easy to see that the general relation between the ranks of these groups in both theories is similar to (\ref{rank-changes}), but without a contribution from the rank of the gauge group coming from the D4-branes between two NS'-branes: 
\beq
\tilde{n}_i = \Delta + n_i   \quad ; \quad \Delta= \displaystyle \sum_{\text{adjacent}} n_i  -  \displaystyle \sum_{\text{dualized, no NS'-NS'}} n_j   \ .  \label{rank-relations-hexagon}
\eeq
In our case $\Delta = 2(N-M)-2N=-2M$.

\bigskip

\subsubsection*{The dimer picture}

\begin{figure}[h]
\begin{center}
\includegraphics[scale=.18]{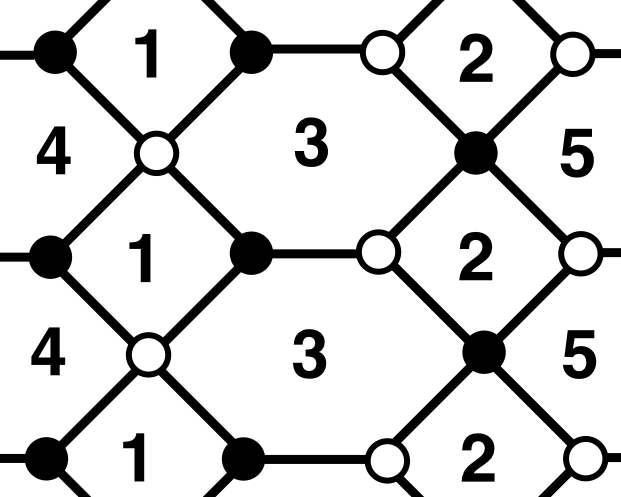}
\caption{\small Dimer describing the theory arising from the brane configuration in figure \ref{tricky-hw2}. }
\label{fig:adj-seiberg-dimer-simple}
\end{center}
\end{figure}

From the above HW brane configuration, we can easliy derive the rules for dualizations involving columns of hexagons in the dimer diagrams of the corresponding orbifold field theories.  The dimer describing the HW configuration in figure \ref{tricky-hw2}(a) is shown in figure \ref{fig:adj-seiberg-dimer-simple}. The hexagon column represents the gauge group between two NS'-branes, while the rhombi columns with labels 1 and 2 bounding the hexagons stand for the two gauge groups between a NS and a NS', and faces with labels 4 and 5 were included since their ranks determine the ranks on the dual gauge theory.  The initial ranks are
\beq
n_1=n_2=n_3=N \quad ; \quad n_4=n_5=N-M \ .
\eeq

The dual theory is also described by this dimer, the only difference being the ranks of the dualized gauge groups. Noting that  we did not include in $\Delta$ is the rank of the hexagon of the dimer, we have that 
\beq
\tilde{n}_i = \Delta + n_i  \quad \text{for } \ i=1, \ 2, \ 3 \quad ; \quad \Delta= \displaystyle \sum_{\text{adjacent}} n_i  -  \displaystyle \sum_{\text{dualized rhombi}} n_j   \ .  \label{rank-relations-hexagon-2}
\eeq
Therefore, the change on the ranks is now given by
\beq
\Delta= (n_4+n_5)-(n_1+n_2)=-2M \ . 
\eeq
Giving the ranks in the dual theory:
\beq
n_1=n_2=n_3=N-2M \quad ; \quad n_4=n_5=N-M \ .
\eeq

\bigskip

\subsubsection*{Adding orientifold points}

\begin{figure}[h]
\begin{center}
\includegraphics[scale=.18]{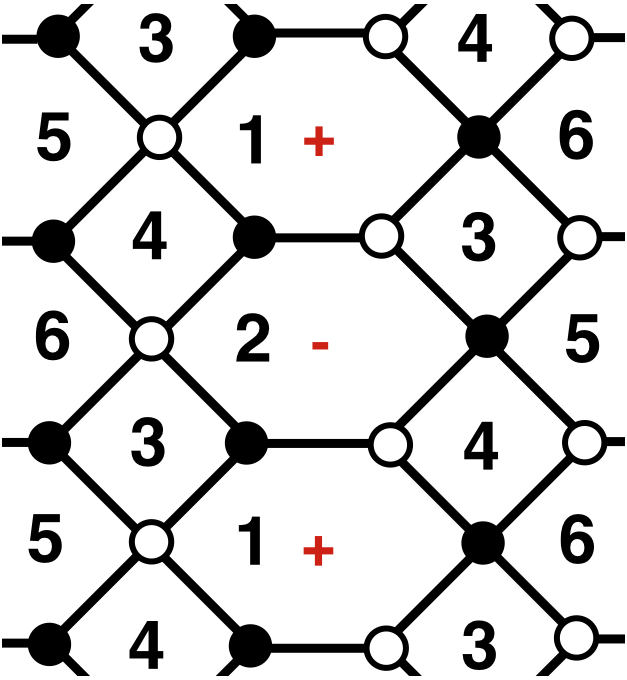}
\caption{\small Dimer describing the $\IZ_2$ orbifold of the one in  figure \ref{fig:adj-seiberg-dimer-simple} with the inclusion of orientifold points. }
\label{fig:adj-seiberg-dimer-orientifold}
\end{center}
\end{figure}

We now deal with the  dualization of hexagons of the dimer found in the main text using the previous knowledge. \fref{fig:adj-seiberg-dimer-orientifold} shows the dimer of interest. Note that it does not admit a HW brane configuration description. It also differs from figure \ref{fig:adj-seiberg-dimer-simple} in that it includes orientifold points projecting $U(N)$ gauge factors down to $SO$ and $USp$ for positive and negative orientifold charge sign. 

To analyze it, we allow for more general ranks of gauge groups compared to the very simple and symmetric choices we picked in the HW picture. We consider a theory with general amounts of branes on each face $n_i$ respecting the orientifold point action. 

The dimer describing the dual theory will also be described by  figure \ref{fig:adj-seiberg-dimer-orientifold} but with different ranks for the dualized gauge groups. Following the ideas in the HW case, the dualization process changes the ranks of all the hexagons and also rhombi on the columns bounding the hexagon column by the single amount $\Delta$. Once again, we compute this amount $\Delta $ using  (\ref{rank-relations-hexagon-2})  and    noting the identification between faces of the dimer due to the orientifold point action. Using this, we find a relation between the ranks given by 
\beq
\tilde{n}_i=n_i +\Delta \quad \text{for } \ i=1, \ 2, \ 3, \ 4  \quad ; \quad \Delta = (n_5 + n_6 )-(n_3 + n_4) \  .
\eeq

\newpage

\bibliographystyle{JHEP}
\bibliography{mybib}

\end{document}